\documentclass[letterpaper,tight,classic,final,pageclassic,orcish,nobm]{iopart}
\usepackage{graphicx,graphicx,epsfig,amsfonts,amssymb}

\usepackage{bm}
\usepackage[all]{xy}

\newcommand{\beq}{\begin{equation}}
\newcommand{\eeq}{\end{equation}}

\newcommand{\qed}{\nobreak \ifvmode \relax \else
      \ifdim\lastskip<1.5em \hskip-\lastskip
      \hskip1.5em plus0em minus0.5em \fi \nobreak
      \vrule height0.75em width0.5em depth0.25em\fi}

\usepackage{bm}
\usepackage[all]{xy}

\begin{document}
\title{Discrete changes of current statistics in periodically driven stochastic systems}

\author {V. Y. {Chernyak}$^{a,b}$ and N.~A. {Sinitsyn}$^{b,c}$ } 

\address{$^a$Department of Chemistry, Wayne State University, 5101 Cass Ave,Detroit, MI 48202}

\address{$^b$ Theoretical Division, Los Alamos National Laboratory, B258, Los Alamos, NM 87545 }

\address{$^c$New Mexico Consortium, Los
  Alamos, NM 87545, USA}

\date{\today}

\begin{abstract}
We demonstrate that the counting statistics of currents in periodically driven ergodic stochastic systems
can show sharp changes of some of its properties in response to continuous changes of the driving protocol.
To describe this effect, we introduce a new topological phase factor in evolution of the moment generating function which is akin to the topological geometric phase in evolution of a periodically driven quantum mechanical system with time-reversal symmetry. This phase leads to the prediction of a sign change for the difference of the probabilities to find even and odd number particles transferred in a stochastic system in response to  cyclic evolution of control parameters. The driving protocols that lead to such a sign change should enclose specific degeneracy points in the space of control parameters. The relation between topology of the paths in the control parameter space and the sign changes can be described in terms of the first Stiefel-Whitney class.
\end{abstract}

\maketitle

\section{Introduction}
\label{sec:intro}

When a system is driven using a periodic protocol, dimensionless currents can be introduced that
describe the fluxes per the driving period, rather than per unit time. The appearance of
stochastic currents in response to  periodic changes of parameters is generally referred to as
the stochastic pump effect \cite{sinitsyn-09review}. Statistical properties of such currents usually change
continuously with changes of the shape of the parameter path. This is, to some degree, expected since usually small changes of kinetic rates in an ergodic Markovian stochastic system, which consists of only a finite number of discrete states, do not lead to sharp changes of the system behavior. Discrete changes of current characteristics can be usually achieved in stochastic systems either with formally infinite number of states or in the limit of low temperature \cite{sinitsyn-09quant, astumian-quantized, shi}
when fluctuations are suppressed and noise cannot smear the behavior near the points of parameter degeneracies. In this Letter, we confront this view and demonstrate that some of the statistical characteristics of currents can depend discontinuously on the choice of the driving protocol, moreover, this property can be found in finite-size ergodic Markov chains.

\section{Moment generating function of currents in a 2-state model}

\begin{figure}
\centerline{\includegraphics[width=2.0in]{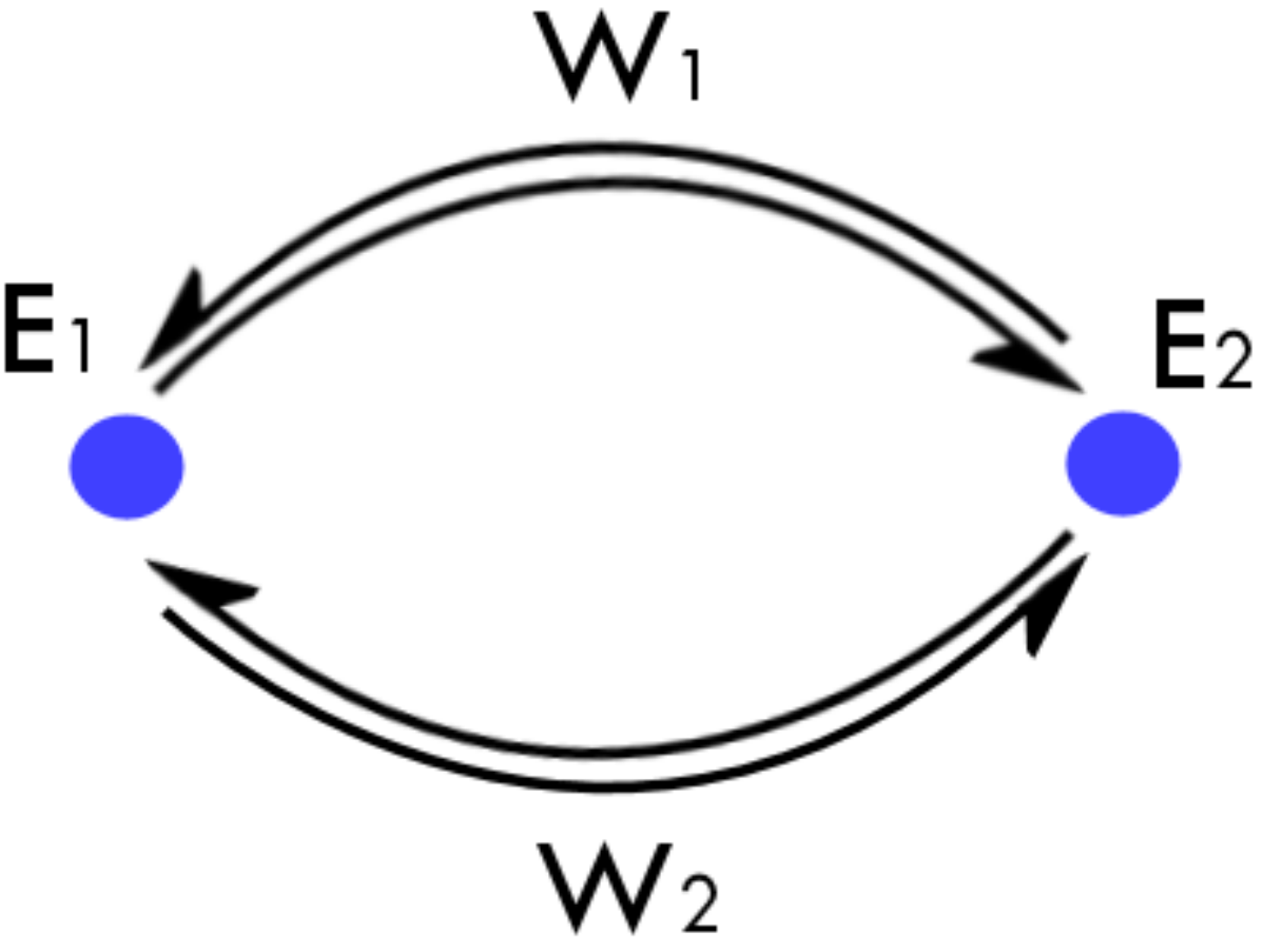}}
  \caption{A model of stochastic transitions between two states along two possible paths. Different paths are
characterized by the values of the corresponding potential barriers $W_1$ and $W_2$. The two states are characterized by
the values of their well depths, $E_1$ and $E_2$. Periodic
modulation of these parameters leads to appearance of a nonzero, on average, circulating current in the preferred (clockwise or counterclockwise) direction. }
\label{2level}
\end{figure}

Consider a simple model illustrated in Fig.~\ref{2level}: a particle can randomly jump between two sites along two paths. Each path can be passed in both directions. We assume the system to be coupled to a heat bath at constant temperature, so that the kinetic rates, $k_{j}^{\alpha}$ of transitions from state $j$ to the other state via the path $\alpha$, with ($j,\alpha=1,2$), satisfy the detailed balance condition and can be parameterized as $k_{j}^{\alpha}=\kappa_{j}g_{\alpha}$, where $\kappa_j=e^{\beta E_{j}}$, $g_{\alpha}=e^{-\beta W_{\alpha}}$, with $\beta =1/(k_BT)$ being the inverse temperature, whereas $E_{j}$ and $W_{\alpha}$ are referred to as the $j$-th potential well energy and the $\alpha$-th path potential barrier height, respectively.

We will study the currents that circulate in a counterclockwise direction, see Fig~\ref{2level}, under the action of a cyclic periodic perturbation. The currents are stochastic and, generally, can be characterized by two sets of probabilities  $ \pi_{j}(n;t)$ that at time $t$ the system is at state $j$, having performed $n$ transitions during the time segment $[0,t]$ through the link $1$ (with the barrier $W_{1}$), counting counterclock/clockwise transition with the $\pm$ sign, respectively. Hereafter, we will notationally suppress the time parameter $t$ in $ \pi_{j}(n;t)$ and other variables.
The Master Equation for $\pi_{j}(n)$ is given by
\begin{equation}
\begin{array}{l}
\dot{\pi}_{1}(n) = -(k_1^1+k_{1}^2)\pi_{1}(n) +k_{2}^1\pi_{2}(n-1)+k_2^2\pi_{2}(n),\\
\\
\dot{\pi}_{2}(n) = -(k_{2}^1+k_{2}^2)\pi_{2}(n) +k_{1}^1\pi_{1}(n+1)+k_{1}^2\pi_{1}(n).
\end{array}
\label{master-2}
\end{equation}
Upon a Fourier transform with respect to $n$, Eq.~(\ref{master-2}) translates into 
\begin{equation}
\dot{{\bm Z}}=\hat{H}(\chi;t){\bm Z}, \;\;\; {\bm Z}=(Z_1,Z_2), \;\;\; Z_{j}(\chi;t)=\sum_{n=-\infty}^{\infty} \pi_{j}(n;t)e^{i\chi n},
\label{master-3}
\end{equation}
with the ``twisted'' evolution operator
\begin{equation}
\hat{H}(\chi)=\left(
\begin{array}{ll}
-\kappa_{1}(g_{1}+g_{2})& \kappa_{2}(g_{1}e^{i\chi}+g_{2})\\
 \kappa_{1}(g_{1}e^{-i\chi}+g_{2}) & -\kappa_{2}(g_{1}+g_{2})
\end{array}\right).
\label{ham2}
\end{equation}

The averaged over the final state MGF $Z(\chi;t)=Z_{1}(\chi;t)+Z_{2}(\chi;t)$ that contains complete information on the current distribution function, including the statistics of rare events, is obtained by solving Eq.~(\ref{master-3}) and can be expressed in terms of a time-ordered exponential \cite{nazarov-03prb}
\begin{equation}
Z(\chi;t)= \langle {\bm 1}|T\exp\left({ \int_{0}^{t}\hat{H}(\chi,t') dt'}\right)|{\bm \pi} \rangle,
\label{pdf2}
\end{equation}
where $\langle {\bm 1}|=(1,1)$, and $|{\bm \pi}\rangle = (\pi_{1},\pi_{2})$ is the vector of the initial system populations.

\section{Degeneracy point of $\hat{H}(\chi)$-eigenvalues}

The lowest current cumulants are described by the properties of $Z(\chi)$ near the $\chi=0$ point, and their behavior in the limit of adiabatically slow evolution of parameters is always defined by the eigenvalue of $\hat{H}(\chi)$ with the largest real part and the corresponding eigenvector \cite{sinitsyn-07epl}.
In the vicinity of  $\chi=0$, the eigenvalues of $\hat{H}(\chi)$ are nondegenerate, provided all kinetic rates in (\ref{ham2}) are nonzero (finite temperature). This follows from an observation that  $\hat{H}(0)$ describes the evolution of probabilities in a 2-state Markov chain \cite{sinitsyn-07epl}. Such an evolution matrix has a unique zero-mode solution corresponding to the unique steady state of an ergodic Markov chain. This uniqueness guarantees that the eigenvalue of $\hat{H}(\chi)$ with the largest real part is nondegenerate for sufficiently small $\chi$.

We will study the effect of geometric phases on the full counting statistics of currents that does not apparently appear when only lowest cumulants of the current distribution are measured, namely, the matrix (\ref{ham2}) has a unique point in the parameter space, where its eigenvalues are degenerate. This happens when simultaneously three conditions $\chi=\pi$, $W_{1}=W_{2}$, and $E_{1}=E_{2}$ are satisfied, which can be seen by explicitly inspecting the eigenvalue problem for the $2\times 2$ matrix given by Eq.~(\ref{ham2}).

Interestingly, the degeneracy conditions do not depend on the inverse temperature $\beta$. The fact that the degeneracy point is encountered at a finite value of the counting parameter,  $\chi=\pi$, means that it influences the properties of the full counting statistics, rather than the lowest cumulants of the current distribution. We note also that at $\chi=\pi$ the  matrix $\hat{H}(\pi)$ has all real entries.

The MGF at $\chi=\pi$,
\begin{equation}
Z(\pi) = P_{0}-P_{1} \equiv  \sum_{j=1;\, k=-\infty}^{2;\,\infty}  \pi_j(2k)- \sum_{j=1; \, k=-\infty}^{2;\,\infty}  \pi_j(2k+1),
\label{zpi}
\end{equation}
has a simple physical meaning, namely, it is the difference of the probabilities $P_{0}$ and $P_{1}$ to make an even and odd number of transitions through a given link, respectively.

The time evolution of $P_{0}-P_{1}$ must be influenced by the presence of the degeneracy point.
The evolution operator $\hat{H}(\pi)$
has two nonpositive eigenvalues
\begin{equation}
\varepsilon_{\pm}=-\kappa_+ g_+ \pm \sqrt{\kappa_+^2g_{-}^2 +\kappa_{-}^2(g_+^2-g_{-}^2)},
\label{egnv}
\end{equation}
with $\kappa_{\pm}=(\kappa_{1}\pm \kappa_{2})/2$, $g_{\pm}=g_{1}\pm g_{2}$. Near their degeneracy point, we parameterize $g_{-}\kappa_{+}=r\sin\phi$ and $\kappa_{-}g_{+}=r\cos\phi$ which yields $\varepsilon_{\pm}\approx-\kappa_{+} g_{+}\pm r$ for the eigenvalues in the limit $g_{-}\ll g_{+}$, $\kappa_{-}\ll \kappa_{+}$.
The eigenvalues as functions of $(r,\phi)$ lie on a two-cone surface. Taking the higher eigenvalue $\varepsilon_{+}$, the eigenstate $|u_{+}(\pi)\rangle \equiv (u_{1},u_{2})$, determined by solving a $2\times 2$ linear problem, has a form $(u_{1},u_{2})=a(\phi)(\sin(\phi/2),-\cos(\phi/2))$
with $a(\phi)$ being an arbitrary periodic function of $\phi$. The requirement on the eigenstate not to be zero imposes that $a(\phi)$ is either strictly positive or strictly negative function, however, the functions $\sin(\phi/2)$ and $\cos(\phi/2)$ change the sign when changing $\phi$ around a full cycle.
If one chooses the smooth field of eigenstates of $\hat{H}(\pi)$ over the space of all possible values of parameters $E_i$ and $W_j$, except the degeneracy point, such a vector field has to be double-valued and, along a nonintersecting path around the degeneracy point, the eigenvectors of $\hat{H}(\pi)$ change the sign.
 To understand, how this effect can influence the stochastic currents, we should look at the evolution of MGF in response to externally induced
periodic changes of kinetic rates in the model.


\section{Evolution of MGF}

\subsection{The case of constant parameters}
When all parameters in the evolution operator are time-independent the time-ordered exponential in (\ref{pdf2}) becomes just a matrix exponent, which can be simplified by introducing the left and right eigenstates $\langle u_{\pm}|$ and $|u_{\pm}\rangle$ of the matrix $\hat{H}_{\pi}$, corresponding to the
eigenvalues in (\ref{egnv}). Then (\ref{pdf2}) gives us
\begin{equation}
P_{0}-P_{1}=C_+e^{\varepsilon_{+}t}+C_{-}e^{\varepsilon_{-}t},  \quad C_{\pm}=\langle {\bm 1} |u_{\pm}(t=0) \rangle \langle u_{\pm}(t=0)|{\bm \pi}\rangle.
\label{stat}
\end{equation}

Since both eigenvalues $\varepsilon_{\pm}$ are negative, the absolute value of $P_{0}-P_{1}$ decays exponentially with time. Moreover, since $\varepsilon_+ > \varepsilon_{-}$, the term $C_{-}e^{\varepsilon_{-}t}$ quickly becomes exponentially suppressed in compared to $C_+e^{\varepsilon_{+}t}$. The coefficients $C_{\pm}$ are totally determined by the initial probability distribution and kinetic rates. Their values describe the initial relaxation processes to the steady state distribution and remain the same even if kinetic rates are changing with time adiabatically \cite{sinitsyn-IET}. At time scales larger than $1/(\varepsilon_+-\varepsilon_{-})$ the sign of $P_{0}-P_{1}$ is determined by the sign of $C_{+}$, which can be found and analyzed explicitly.
Depending on the initial choice of parameters, including initial state probabilities, $C_{+}$ can be either negative or positive, however, the sign of $P_{0}-P_{1}$ is fixed after the term with coefficient $C_{-}$ becomes exponentially suppressed. In other words, the sign of $P_{0}-P_{1}$ is the invariant of the evolution which can take values in a discrete two-state set. 

\subsection{Geometric phase}
We further ask a question whether the sign of $P_{0}-P_{1}$ can be switched by  adiabatically slow evolution of parameters $g_{-}$ and $\kappa_{-}$ around a cycle. In the adiabatic limit, a naive strict quasi-steady-state approximation predicts that a slow perturbation merely modulates the decay exponents in (\ref{stat}) and the vector ${\bm Z}(\pi)$ is given by instantaneous eigenvector $|u_{+}\rangle$. This means that, in principle, the sign of $P_{0}-P_{1}$ can change during the evolution since the components
of $|u_{+} \rangle$  change with time and their sum can switch the sign. However, if the parameters complete a cycle, the initial and final eigenstates  coincide, while the coefficient $C_+$ is time-independent. Hence, one may not expect to find a sign change of   $P_{0}-P_{1}$ at the end of the cycle. This analysis, however, does not include the effect of geometric phases, which can be non-perturbative \cite{mostafazadeh-08}.

First, consider the case when a cyclic path in the space of control parameters does not enclose the point $g_{-}=\kappa_{-}=0$. We can always choose a smooth parametrization of the eigenvectors of $\hat{H}(\pi)$ in the region $S_{\bm C}$ inside the contour.
\begin{figure}
\centerline{\includegraphics[width=2.8in]{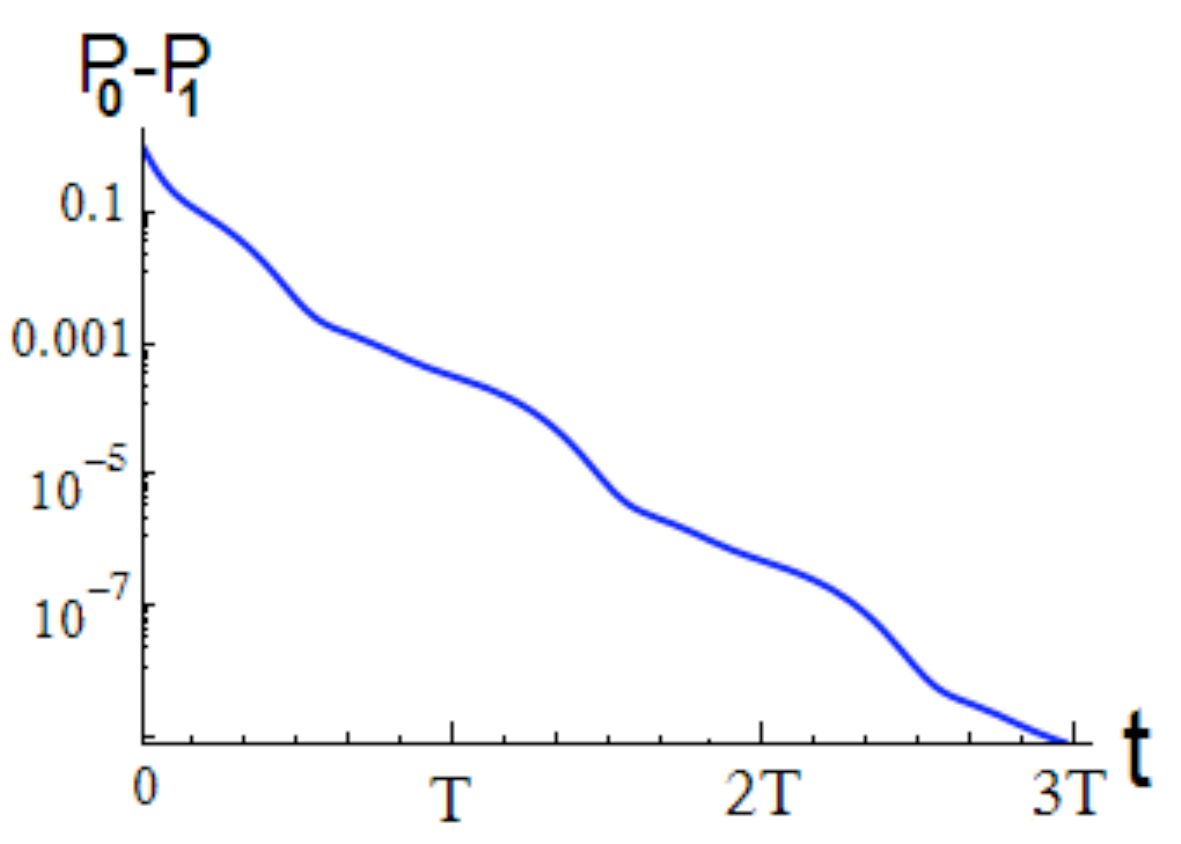}}
  \caption{Evolution of the parity probability difference in a driven 2-state stochastic
system. The choice of parameters is $\kappa_{+}=2$, $g_{+}=1$,
 $\kappa_{-}=0.8+2\cos(2\pi t/T)/3$,  $g_{-}=0.4+\sin(2\pi t/T)/3$, $T=30$.
The contour in the space of control parameters does not enclose
the degeneracy point.}
\label{z2-smooth}
\end{figure}
According to \cite{sinitsyn-09review,sinitsyn-07epl,ohkubo-08stat,sinitsyn-07prl,hanggi-10}, $P_{0}-P_{1}$, after periodic and adiabatically slow evolution of the parameters around the contour ${\bm C}$ during time $T$, is given by
\begin{equation}
P_{0}-P_{1}=C_{+}\exp\left(\int_{0}^{T}\varepsilon_{+}(t)dt + \varphi_{\rm geom}({\bm C})\right),
\label{geometric}
\end{equation}
where  the geometric contribution to the exponent in (\ref{geometric}) is given by
\begin{equation}
\varphi_{\rm geom}({\bm C})=\int_{S_{\bm C}} dg_{-}d\kappa_{-} F_{g_{-},\kappa_{-}},
\label{curvature}
\end{equation}
where $F_{g_{-},\kappa_{-}}=\langle \partial_{g_{-}} u_{+}  |\partial_{\kappa_{-}} u_{+}  \rangle -
\langle \partial_{\kappa_{-}} u_{+} |\partial_{g_{-}} u_{+} \rangle$
or explicitly
\begin{equation}
F_{g_{-},\kappa_{-}}
=-\frac{\kappa_{-}g_{-}\kappa_+g_+}{2(\kappa_{+}^2g_{-}^2+\kappa_{-}^2(g_+^2-g_{-}^2))^{3/2}}.
\label{ff}
\end{equation}

The eigenvalues $\varepsilon_{+}(t)$ and the geometric phase $\varphi_{\rm geom}({\bm C})$ in (\ref{geometric}) are both real, and therefore the exponential term does not affect the sign of $P_{0}-P_{1}$. Fig.~\ref{z2-smooth} shows the results of our numerical solution of the evolution equation (\ref{master-3}) at $\chi=\pi$. The fact that $P_{0}-P_{1}$ has the same sign during the whole evolution is not generic. Generally, it can change but the fact that it does not change after each period of driving protocol is always confirmed if the contour does not enclose the degeneracy point. In other words, the sign of $P_{0}-P_{1}$ is an invariant of an adiabatic change of parameters along a closed contour that does not enclose the degeneracy point.
\begin{figure}
\centerline{\includegraphics[width=2.8in]{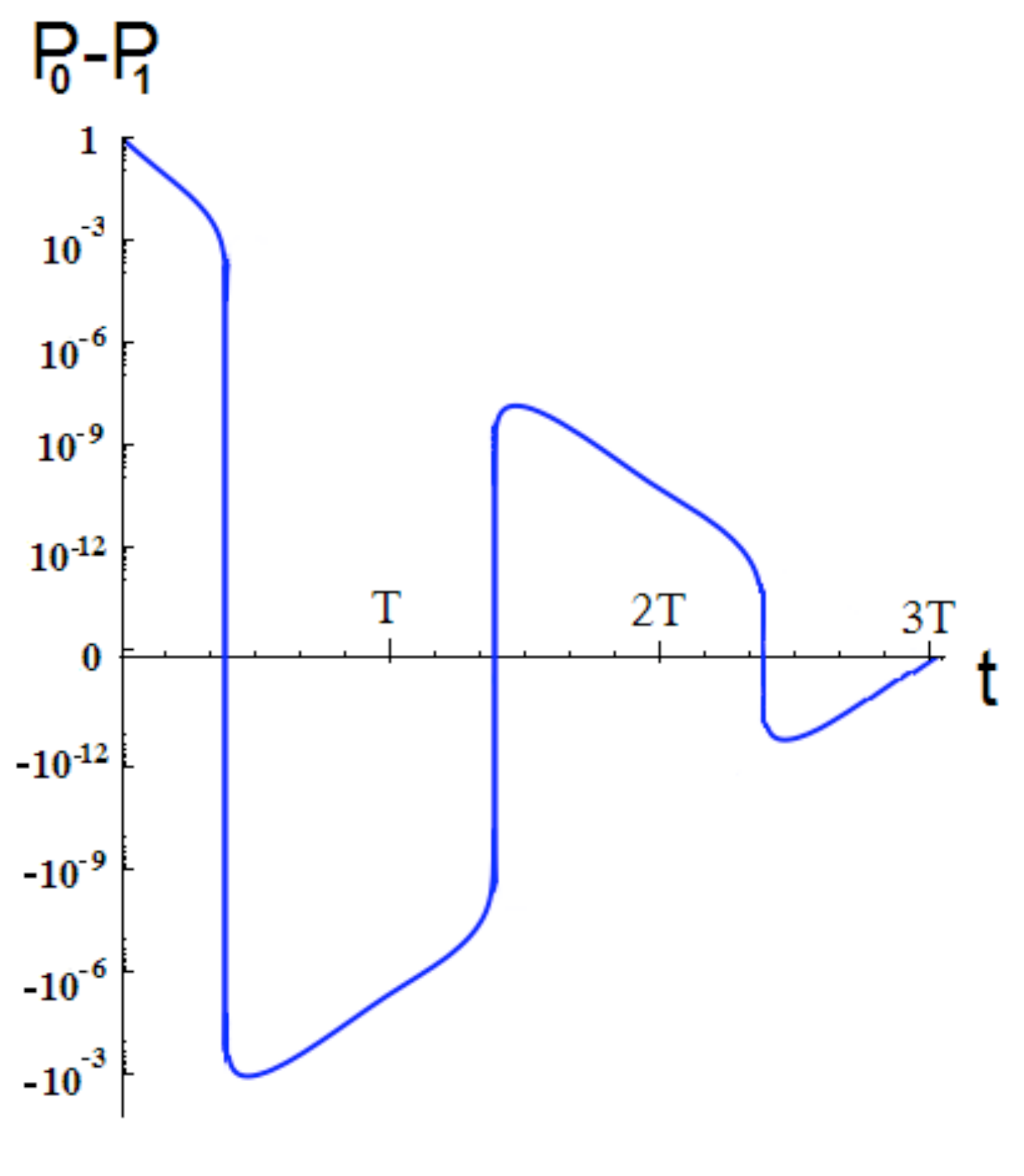}}
  \caption{Evolution of the probability difference $P_{0}-P_{1}$ in a driven 2-state stochastic
system when a contour in the space of control parameters  encloses
the degeneracy point. The choice of parameters is $\kappa_{+}=1$, $g_{+}=1$,
 $\kappa_{-}=2\cos(2\pi t/T)/3$,  $g_{-}=\sin(2\pi t/T)/3$, $T=30$.
}
\label{z2-fin}
\end{figure}

\subsection{Robust switching of the sign of  $P_{0}-P_{1}$}
We now consider the case when the contour is cyclic and encloses the degeneracy point $g_{-}=\kappa_{-}=0$.  The previous arguments cannot be applied directly since at the degeneracy
point the adiabatic approximation breaks down and the parametrization of the eigenstates inside the contour cannot be chosen smoothly. They do, however, imply that ${\rm sgn}(P_{0}-P_{1})$
does not change upon deformations of ${\bm C}$ that do not encounter the degeneracy point reducing our analysis to the contours, restricted to a small neighborhood of the degeneracy point. Our analysis, in section 3, of the vicinity of the crossing point showed that the smooth choice of eigenstates along a contour requires that, upon completion of the cycle, the eigenvector changes sign.
Physically, this change of sign means that the probability difference $P_{0}-P_{1}$ changes the sign. Our numerical solution of Eqs.~(\ref{master-3}) for evolution along a contour that encloses the degeneracy point, is shown in Fig.~\ref{z2-fin}. It confirms the sign change prediction. Note that this result is nonperturbative. It cannot appear as an effect of small non-adiabatic corrections to the decaying exponent in (\ref{geometric}). The factor $-1$ is a topological factor rather than a geometric phase effect in a sense of Ref.~\cite{sinitsyn-07epl}. One can see from Fig.~\ref{z2-fin} that this factor does appear after an adiabatic cyclic evolution of the control parameters that enclose the degeneracy point. More generally this implies that ${\rm sgn}(P_{0}-P_{1})=(-1)^{n({\bm C})}{\rm sgn}C_{+}$ with $n({\bm C})$ being the number of times the cycle ${\bm C}$ winds around the degeneracy point. This property provides an approach to manipulate this sign with high robustness against perturbations of a driving protocol.

\section{Topological interpretation of the invariant and generalizations}
The sign of the probability difference in (\ref{zpi}) is, in fact, a topological invariant of the driving protocol, the latter being viewed as a cycle in the space $M\cong \mathbb{R}^{2}\setminus\{0\}$ of parameters $x=(g_{-},\kappa_{-})$ where degeneracy of $\hat{H}(\pi;x)$ never occurs. By associating with $x\in M$ the ground state of $\hat{H}(\pi;x)$, the former defined up to a real multiplicative factor, we build a linear real fiber bundle $E$ over $M$, or equivalently  a map $g:M\to \mathbb{RP}^{1}$. Here the projective space $\mathbb{RP}^{1}$ is interpreted as a circle $S^{1}$ of ``normalized states'' $(u_{1})^{2}+(u_{2})^{2}=1$ with the opposite points on the circle identified, since there is no preferred way to define the sign of the ground state. The map $g$ associates with a driving protocol ${\bm C}$ a cycle $g({\bm C})$ in $\mathbb{RP}^{1}\cong S^{1}$, and each time ${\bm C}$ winds around the degeneracy point, $g({\bm C})$ produces a single rotation in $\mathbb{RP}^{1}$, so that $n({\bm C})=n(g({\bm C}))$. The last statement can be reformulated in terms of the first Stiefel-Whitney class $c_{1}=c_{1}(E)$ \cite{milnor-stasheff-74} of $E$ that can be viewed as a $\mathbb{Z}_{2}$-valued function on cycles in $M$, so that $c_{1}({\bm C})=n(g({\bm C}))\;{\rm mod}\;2$ and the invariant $(-1)^{n({\bm C})}=(-1)^{c_{1}({\bm C})}$ of the counting statistics is obtained by evaluating the Stiefel-Whitney class at the driving protocol.

In this letter we studied in some detail a $\mathbb{Z}_{2}$-invariant in a simple $2$-node system. In a general network represented by a finite connected graph there is a set of invariants, we are going to describe briefly here, with the analysis of the general case postponed to a further publication. With any ``independent'' set ${\bm\alpha}=\{\alpha_{1},\ldots,\alpha_{k}\}$ of links (i.e., if they are all withdrawn the network still remains connected) we can associate a $\mathbb{Z}_{2}$-valued observable $I_{{\bm\alpha}}({\bm C})={\rm sgn}\sum_{{\bm a}}(-1)^{\sum_{j}a_{j}}P_{{\bm a}}({\bm\alpha};{\bm C})$, where ${\bm a}=\{a_{1},\ldots,a_{k}\}$ is a set of binary variables, and $P_{{\bm a}}({\bm\alpha};{\bm C})$ is the joint probability that for protocol ${\bm C}$ the total number of jumps over link $\alpha_{j}$ was even/odd for $a_{j}=0,1$ respectively. The protocol is assumed to go through the space $M_{{\bm\alpha}}$ of parameters that avoid the ground-state degeneracy of the twisted operator $\hat{H}({\bm\chi})$, with $\chi_{\gamma}=\pi$ for chosen links $\gamma\in{\bm\alpha}$, and $\chi_{\gamma}=0$, otherwise. Similar to the simple two-node case the ground state defines a linear bundle $E_{{\bm\alpha}}$ over $M_{{\bm\alpha}}$ with the first Stiefel-Whitney class $c_{1}=c_{1}(E_{{\bm\alpha}})$, or equivalently a map $g:M_{{\bm\alpha}}\to \mathbb{RP}^{N-1}$ with $N$ being the number of nodes in the network. Viewing $\mathbb{RP}^{N-1}$ as an $(N-1)$-sphere with opposite points identified we have a two-fold cover $S^{N-1}\to\mathbb{RP}^{N-1}$. Since all cycles in $S^{N-1}$ are contractible, there are two equivalence classes of cycles ${\bm s}$ in the projective space: the ones produced by cycles in $S^{N-1}$ and the ones produced by contours that connect opposite points in $S^{N-1}$; this is referred to as $[{\bm s}]=0$ and $[{\bm s}]=1$, respectively. The topological factor associated with a driving protocol is given by $(-1)^{[g({\bm C})]}=(-1)^{c_{1}({\bm C})}$ so that the counting-statistics invariant $I_{{\bm\alpha}}({\bm C})=(-1)^{c_{1}({\bm C})}{\rm sgn}C_{+}$, with $C_{+}$ depending on the initial distribution only, being independent of the driving protocol.

The topological factor is very similar to the topological Berry phase in quantum mechanics of systems with time-reversal symmetry (no magnetic fields), where $\hat{H}(x)$ is a family of Hamiltonians, represented by real Hermitian operators \cite{topological-phase,mostafazadeh-08}. In our case the operators are real, however, apparently non-Hermitian. They are in fact Hermitian with respect to a scalar product that depends on the kinetic rates. The latter results in the geometric contribution $\varphi_{{\rm geom}}({\bm C})$ in Eq.~(\ref{geometric}) that does not affect ${\rm sgn}(P_{0}-P_{1})$.

\section{Discussion}
We demonstrated that some of the properties of the full counting statistics of currents in finite size stochastic models can be controlled by periodic changes of the control parameters without
continuous dependence of the result of operation on the choice of the path in the parameter space.
This finding reveals the importance of the degeneracy points of the eigenvalues of $\hat{H}(\chi)$, governing the evolution of the current MGF.
Although the effect is the property of the full counting statistics \cite{nazarov-03prb,jarzysnki-97prl,bochkov-77,crooks-98, moskalets-review,kamenev-04pre}, it is measurable experimentally, since the probabilities of rare events in the considered model are suppressed at least exponentially. As a result, only a finite number of terms in the definition of the parity difference (\ref{zpi}) should be determined, which is achievable by repeating the measurement of the number of transition through a link after sufficiently many cycles.
The simplicity of our model suggests that the topological geometric factors could be a
generic property of current statistics. The phenomena that they represent in general situations of multistate or Langevin evolution can be much more complex than the sign switching that we demonstrated.

\section*{Acknowledgment}
We a grateful to John R Klein for useful discussions and comments.
{\it  This material is based upon work supported by NSF under Grants No. CHE-0808910, ECCS-0925365, and in part by DOE under Contract No.\ DE-AC52-06NA25396.}

\end{document}